# Solution of 3D magnetization problems for superconducting film stacks

**Leonid Prigozhin[1] and Vladimir Sokolovsky[2]**

[1]J. Blaustein Institutes for Desert Research, Ben-Gurion University of the Negev, Sede Boqer Campus 84990, Israel

[2]Physics Department, Ben-Gurion University of the Negev, Beer-Sheva 84105, Israel

E-mail: leonid@math.bgu.ac.il and sokolovv@bgu.ac.il

**Abstract**
A stack of coated conductors is a perspective configuration for various applications of high temperature superconductors. We present an efficient fast Fourier transform-based numerical method for magnetization problems for a stack of flat films of the same (arbitrary) shape and compare it with the recently proposed finite element methods. For stacks containing a large number of densely packed films an accurate solution can be obtained as a properly rescaled solution for a stack of only several films. For an infinite stack the problem simplifies and becomes similar to that for a single film.

## 1. Introduction

Progress in fabrication and commercial availability of high temperature coated conductors made stacks of such conductors an attractive alternative to bulk superconductors for trapping strong magnetic fields, magnetic levitation, etc. Advantages of coated conductors are their high critical current density, better mechanical and thermal stability due to the metal substrate layers, higher degree of uniformity, and flexibility in adaptation to different configurations (see, e.g., [1-8] and the references therein). Stacks of coated conductors experience much weaker crossed field demagnetization (see [9, 10]); this property is especially important for the superconducting magnets in electric motors.

Typically, the width to thickness ratio of the superconducting layer in coated conductors is between 1000 and 10000, which justifies modeling using the infinitely thin film approximation. For stacks of a few infinitely long strips, the two-dimensional (2D) magnetization problems were solved numerically in, e.g., [11-13]. Efficient solution for a densely packed stack of many such strips can be obtained using the anisotropic bulk model proposed by Clem, Claassen and Mawatari [14], see also [15, 16]. For an infinitely high stack of equally spaced strips obeying the Bean critical-state model an analytical solution was found by Mawatari [17]; this solution is an algebraic transformation of the well-known solution for a single strip [18]. Recently, a three-dimensional (3D) problem, magnetization of a stack of square films (the stack benchmark problem), was solved in the anisotropic bulk approximation using several finite element methods [19, 20].

Here we present a new numerical method for 3D magnetization problems for a finite stack of flat films of the same (arbitrary) shape. The method uses the fast Fourier transform (FFT) for approximation in space and the method of lines for integration in time. It is an extension of the numerical method for thin film magnetization problems proposed by Vestgården et al. [21, 22] and modified in our work [23]. The outline of this paper is as follows. First, we formulate the problem and describe our method (Section 2), then solve the stack benchmark problem and compare our approach with those in [19, 20] (Section 3). We also consider the limit of high stacks (Section 4), show that for an infinite stack of films the problem is simplified and becomes similar to that for a single film of the same shape, and solve it numerically. Finally (Section 5), our results are discussed.

## 2. Stack magnetization
*2.1 Mathematical formulation*

We consider a stack of $N$ thin equidistant superconducting films, $\{(x, y, z_m) : (x, y) \in \Omega, z_m = md\}$, where $m = 1, ..., N$ and $d$ is the distance between films. For simplicity, we assume that the open domain $\Omega \subset R^2$ is simply connected, the normal to films z-component of the applied magnetic field is uniform, $h_{e,z} = h_{e,z}(t)$, and in all films the same power law,

$$e_m = e_c \left( \frac{|j_m|}{j_c} \right)^{n-1} \frac{j_m}{j_c}, \qquad (1)$$

holds for the sheet current densities, $j_m$, and the parallel to films electric field components, $e_m$. Here $e_c = 10^{-4}$ V/m, the sheet critical current density $j_c$ and the power $n$ are assumed constant. Since $\nabla \cdot j_m = 0$ in $\Omega$ and the normal component of $j_m$ on the domain boundary $\Gamma$ is zero, there exist stream functions $g_m$ such that

$$j_m = \bar{\nabla} \times g_m \qquad (2)$$

(i.e. $j_{m,x} = \partial_y g_m$, $j_{m,y} = -\partial_x g_m$) and $g_m |_\Gamma = 0$. Although our method is, in general, similar to that for a single film in [21, 23], a special consideration is needed to express the functions $g_m$ via the z-component of magnetic field. Let us extend $g_m$ by zero to the outer domain $\Omega_{out} = R^2 \setminus (\Omega \bigcup \Gamma)$ and define $G_{m-l}(r) = \left( 4\pi \sqrt{r^2 + d^2(m-l)^2} \right)^{-1}$, where $r = (x, y)$ and $r = |r|$. By the Biot-Savart law

$$h_{m,z} - h_{e,z} = \sum_{l=1}^{N} \nabla \times \int_\Omega G_{m-l}(r - r') j_l(r', t) dr' =$$

$$\sum_{l=1}^{N} \bar{\nabla} \times G_{m-l} * \bar{\nabla} \times g_l = -\sum_{l=1}^{N} \left( \partial_x G_{m-l} * \partial_x g_l + \partial_y G_{m-l} * \partial_y g_l \right),$$

where $\nabla \times u = \partial_x u_y - \partial_y u_x$ is the scalar 2D curl of a vector function, $h_{m,z}$ is the z-component of magnetic field on the $m$-th film, and * denotes the 2D convolution. Applying the Fourier transform $F[f] = \int_{R^2} f(r) e^{-ikr} dr$ we obtain $F[h_{m,z} - h_{e,z}] = k^2 \sum_{l=1}^{N} F[G_{m-l}] F[g_l]$, where $k^2 = k_x^2 + k_y^2$. For $k \neq 0$ this yields a linear algebraic system for $F[g_l]$, $l = 1, ..., N$:

$$\sum_{l=1}^{N} F[G_{m-l}] F[g_l] = \frac{1}{k^2} F[h_{m,z} - h_{e,z}]. \qquad (3)$$

Since $G_{m-l}(r)$ is radially symmetric, its 2D Fourier transform is ([24], Appendix I)

$$F[G_{m-l}](k) = 2\pi \int_0^\infty J_0(kr) G_{m-l}(r) r \, dr = \frac{1}{2} \int_0^\infty \frac{J_0(kr) r}{\sqrt{r^2 + d^2(m-l)^2}} \, dr,$$

where $J_0$ is the Bessel function of the first kind of zero order. Furthermore, using [25], formula 11.4.16 for $l = m$ and formulas 11.4.44 and 10.2.17 for $l \neq m$, we find $F[G_{m-l}](k) = (2k)^{-1} e^{-kd|m-l|}$ and rewrite the linear system (3) as

$$\sum_{l=1}^{N} A_{m,l}(k) F[g_l](k) = \frac{2}{k} F[h_{m,z} - h_{e,z}](k), \tag{4}$$

where $A_{m,l}(k) = q^{|m-l|}$ with $q(k) = e^{-kd}$. For $k \neq 0$ the symmetric $N \times N$ matrix $\mathbf{A}$ has a simple three-diagonal inverse:

$$\mathbf{A}^{-1}(k) = \frac{1}{1-q^2} \begin{pmatrix} 1 & -q & 0 & \cdots & & 0 \\ -q & q^2+1 & -q & 0 & \cdots & 0 \\ 0 & -q & q^2+1 & -q & \cdots & 0 \\ & & \cdots & & & \\ 0 & \cdots & 0 & -q & q^2+1 & -q \\ 0 & \cdots & & 0 & -q & 1 \end{pmatrix}.$$

Hence, if the functions $F[h_m - h_{e,z}]$ are known, the functions $F[g_m]$ remain undetermined by (4) only for $k = 0$. These values correspond to additive constants in the real space and we express $\mathbf{g} = (g_1, \ldots, g_N)^T$ via $\delta \mathbf{h}_z = (h_{1,z} - h_{e,z}, \ldots, h_{N,z} - h_{e,z})^T$ as

$$\mathbf{g} = \mathbf{\Phi}[\delta \mathbf{h}_z] := F^{-1}\left[\frac{2}{k} \mathbf{A}^{-1}(k) F[\delta \mathbf{h}_z]\right] - \mathbf{C}(t) \tag{5}$$

where $\frac{2}{k} \mathbf{A}^{-1}(k)$ is replaced by the zero matrix for $k = 0$, $\mathbf{C} = (C_1, \ldots, C_N)^T$, and the shifts $C_m(t)$ are chosen to satisfy the conditions $\int_{\Omega_{\text{out}}} g_m d\mathbf{r} = 0$ at each moment in time. Differentiating (5), we obtain

$$\dot{\mathbf{g}} = \mathbf{\Phi}[\delta \dot{\mathbf{h}}_z] = F^{-1}\left[\frac{2}{k} \mathbf{A}^{-1}(k) F[\delta \dot{\mathbf{h}}_z]\right] - \dot{\mathbf{C}} \tag{6}$$

with $\dot{\mathbf{C}}$ such that $\int_{\Omega_{\text{out}}} \dot{\mathbf{g}} d\mathbf{r} = \mathbf{0}$; here and below the dot means differentiation in respect to time. Following [21, 23], we can now formulate an evolutionary problem for $\mathbf{g}$. Let all stream functions $g_m$ be known at time $t$. Then we also know the sheet current densities $\mathbf{j}_m = \bar{\nabla} \times g_m$ and can find the parallel to films electric field components $\mathbf{e}_m = \rho(j_m) \mathbf{j}_m$ in each film using (1). By the Faraday law we have

$$\dot{h}_{m,z} = -\mu_0^{-1} \nabla \times \mathbf{e}_m = \mu_0^{-1} \nabla \cdot [\rho(|\nabla g_m|) \nabla g_m] \quad \text{in} \quad \Omega, \tag{7}$$

where $\mu_0$ is the magnetic permeability of vacuum. This does not determine the evolutionary problem (6) yet because the electric fields $\mathbf{e}_m$ and, hence, $\dot{h}_{m,z}$ and $\delta \dot{\mathbf{h}}_z$ in $\Omega_{\text{out}}$ remain unknown. However, in the outer domain $\mathbf{g}$ should remain zero, which is an implicit condition for $\delta \dot{\boldsymbol{h}}_z |_{\Omega_{\text{out}}}$: equation (6) should hold with $\dot{\boldsymbol{g}}|_{\Omega_{\text{out}}} = \boldsymbol{0}$. This condition can be resolved iteratively.

*2.2. Iterations*

To find the derivative $\dot{\boldsymbol{g}}$ for a given $\boldsymbol{g}$, for each film we find $\boldsymbol{j}_m = \bar{\nabla} \times \boldsymbol{g}_m$ and set

$$\boldsymbol{e}_m = \begin{cases} \rho(|\boldsymbol{j}_m|) \boldsymbol{j}_m & \text{in } \Omega, \\ \rho_{\text{out}} \boldsymbol{j}_m & \text{in } \Omega_{\text{out}} \end{cases}$$

with a sufficiently high constant fictitious resistivity $\rho_{\text{out}}$. Then we define $\dot{h}_{m,z}^{(0)} = -\mu_0^{-1} \nabla \times \boldsymbol{e}_m$, $\delta \dot{h}_{m,z}^{(0)} = \dot{h}_{m,z}^{(0)} - \dot{h}_{e,z}$, and set an initial approximation $\dot{\boldsymbol{g}}^{(0)} = \boldsymbol{\Phi}[\delta \dot{\boldsymbol{h}}_z^{(0)}]$. It is desirable to improve this approximation: if the resistivity $\rho_{\text{out}}$ is high, the arising evolutionary problem is stiff and its integration inhibited, otherwise a non-negligible current in the non-conducting domain appears. Hence, on the *i*-th iteration, we improve $\dot{h}_{m,z}^{(i)}$ by subtracting the time derivative of the field induced at $z = md$ by the stray current outside the film. For normal to the *m*-th film component of this field the derivative can be presented as $F^{-1}\left[(k/2) F[\dot{g}_{m,\text{out}}]\right]$, where

$$\dot{g}_{m,\text{out}} = \begin{cases} 0 & \text{in } \Omega, \\ \dot{g}_m & \text{in } \Omega_{\text{out}}. \end{cases}$$

Since values of $\dot{h}_{m,z}$ in the film itself are determined by (7), we update $\dot{h}_{m,z}$ only in $\Omega_{\text{out}}$ and set, for all $m$,

$$\dot{h}_{m,z}^{(i+1)}|_{\Omega_{\text{out}}} = \dot{h}_{m,z}^{(i)}|_{\Omega_{\text{out}}} - \lambda F^{-1}\left[(k/2) F[\dot{g}_{m,\text{out}}^{(i)}]\right]\Big|_{\Omega_{\text{out}}} - A_m^{(i)}, \tag{8}$$

where the shifts $A_m^{(i)}$ ensure the conditions $\int_{R^2} \left(\dot{h}_{m,z}^{(i+1)} - \dot{h}_{e,z}\right) d\boldsymbol{r} = 0$ and $\lambda$ is a relaxation parameter, then find

$$\dot{\boldsymbol{g}}^{(i+1)} = \boldsymbol{\Phi}[\delta \dot{\boldsymbol{h}}_z^{(i+1)}]. \tag{9}$$

Provided these iterations converge, $\dot{g}_{m,\text{out}}^{(i)}$ tend to zero for all $m$ as desired. In practice, the iterations (8),(9) are performed on a finite grid and eliminate the stray currents in each film plane except narrow boundary layers outside the films; the width of these layers scales with the grid step size. The fictitious resistivity $\rho_{\text{out}}$ should be sufficient to suppress currents only in these layers and, usually, does not need to be very high.

*2.3. Implementation details*

Our computer implementation of the described iterative method is similar to that in [23] for a single film. For spatial discretization we use a regular $N_x \times N_y$ grid in a rectangular computation domain $D$ containing the film domain $\Omega$ and several times larger. Values of all variables are defined in the grid nodes; the continuous Fourier transform and its inverse are replaced by their discrete analogues on this grid and computed using the FFT algorithm. The 2D curl operators $\bar{\nabla} \times \boldsymbol{g}_m$ and $\nabla \times \boldsymbol{e}_m$ are computed in the Fourier space with the Gaussian smoothing to suppress high frequency oscillations. The parameter of smoothing should be of the order of grid cell size; in our simulations it was taken equal to the grid cell diagonal. The same Gaussian smoothing is applied to (8).

We used Matlab R2017a, its standard FFT software, and the ordinary differential equation solver ode23 (with the relative tolerance $2 \cdot 10^{-4}$) to integrate the spatially discretized problem in time. All simulations were performed on a computer with the Intel® Core™ i7-4770 CPU @ 3.40 GHz processor, 16 GB RAM, 64-bit Windows 10.

## 3. The benchmark problem

Employed in applications are often stacks of a large number of densely packed coated conductors with the distance between the superconducting films much less than the film size. In such cases efficient solution of magnetization problems can be obtained using homogenization and transition to the anisotropic bulk model [14-16]. The stack of $N$ films with the neighboring film distance $d$ and the current-voltage relation (1) is replaced by a cylindrical bulk superconductor of the height $H = Nd$ and cross-section $\Omega$, characterized by the infinite resistivity in the $z$-axis direction and the power current-voltage law

$$\boldsymbol{e} = e_c \left( \frac{|\boldsymbol{J}|}{J_c} \right)^{n-1} \frac{\boldsymbol{J}}{J_c} \qquad (10)$$

for the parallel to $xy$-plane component $\boldsymbol{e}$ of the electric field and the bulk current density $\boldsymbol{J}$. Here the critical current density $J_c = j_c / d$ equals the film sheet critical current density averaged over the layer of thickness $d$.

For the fixed parameters $J_c$ and $H$ the anisotropic bulk model solution is the limit of solutions to stack problems with the sheet critical current densities $j_c^N = J_c d_N = J_c H / N$ as the number of films $N$ tends to infinity. Using different finite element methods, magnetization of such initially non-magnetized anisotropic bulk $10 \times 10 \times 1$ mm$^3$ superconductor was simulated in [19, 20] for the critical current density $J_c = 10^8$ Am$^{-2}$, the power $n = 25$, and the sinusoidal applied field $h_{e,z}$ with the amplitude 100 mT and frequency 50 Hz. Presented in these works are, in particular, the current density distribution at the first peak of the applied field, $t = 0.25T$, where $T$ is the period, and the loss, $Q_{mh} = -\mu_0 \oint m_z \mathrm{d} h_{e,z}$, computed for the cycle $0.25T \leq t \leq 1.25T$; here $m_z$ is the $z$-component of the magnetic moment.

One of the methods, MEMEP [19], was based on a variational formulation for the effective magnetization; the algorithm, written in C++, was parallelized (see [26]) to accelerate time consuming computations. For a similar spatial resolution the problem was solved in [19] also using the popular $h$-formulation and COMSOL Multiphysics. The $h$-formulation was employed also in [20], where an advanced highly parallelized finite element algorithm was realized in the open source simulation software FEMPAR. For personal computers similar to ours, solution of the benchmark problem on a $71 \times 71 \times 7$ finite element mesh by the MEMEP method took 6.0 days; COMSOL needed 1.7 days to solve the benchmark problem using a similar mesh inside the superconductor and the $h$-formulation-based method (see [19]). Computed by these two methods, losses per cycle, $Q_{mh}$, were, respectively, 3.50 and 3.45 mJ. In [20], the computation time for this problem was not reported; the $Q_{mh}$ value was 3.46 mJ.

Our attempt to solve the stack benchmark problem using the 3D FFT-based method for bulk superconductors [23, 27] was unsuccessful: it was difficult to obtain an accurate solution with resistivity in the $z$-axis direction fully suppressing the corresponding current density component. Much better result was obtained using a different approach, suggested for two-dimensional stack problems in [16]. As was noted there, if the ratio of the film distance $d$ to the half of superconducting strip width is less than 0.05, the difference between the AC losses in a stack and the corresponding anisotropic bulk superconductor does not exceed 2%. Hence, instead of the full homogenization and transition to the anisotropic bulk model, it is possible to replace the densely packed $N$-film stack by a stack with a lower number of tapes, $N_0 \ll N$, the same height $H = dN = d_0 N_0$, and the new sheet critical current density $j_{c,0} = d_0 J_c = d_0 j_c / d$.

For the stack benchmark problem, the ratio of stack height to the film half-side $a$ is 0.2. With only 4 films we already have $d/a = 0.05$ and so even the 4-film stack should be a good approximation to the benchmark problem. Since the normal to film component of the magnetic moment can be conveniently expressed via the stream function, $m_z = \frac{1}{2}\int_\Omega \boldsymbol{r} \times \boldsymbol{j} \mathrm{d}\boldsymbol{r} = \int_\Omega g \mathrm{d}\boldsymbol{r}$, the AC loss per period for the film stacks was calculated as

$$Q_{mh} = -\int_{0.25T}^{1.25T} \dot{h}_{e,z} \left( \sum_{l=1}^{N} \int_\Omega g_l \mathrm{d}\Omega \right) \mathrm{d}t.$$

In simulations performed for the 4- and 6-film stacks we used dimensionless variables,

$$(x', y', z') = \frac{(x, y, z)}{a}, \quad t' = \frac{t}{t_0}, \quad \boldsymbol{e}' = \frac{\boldsymbol{e}}{e_c},$$

$$\boldsymbol{J}' = \frac{\boldsymbol{J}}{J_c}, \quad \boldsymbol{h}' = \frac{\boldsymbol{h}}{J_c a},$$

where $a = 5 \cdot 10^{-3}$ m and $t_0 = \mu_0 J_c a^2 / e_c = 31.5$ s. The dimensionless applied field was $h'_{e,z} = 0.1587 \sin(9896 t')$ and its period $T' = 6.349 \cdot 10^{-4}$. The uniform $512 \times 512$ grid was defined in the computation domain $D$; this domain was about five times larger than the film domain $\Omega$ containing $101 \times 101$ grid nodes. The iterations (8)-(9) were performed with the relaxation parameter $\lambda = d'/2$ until the average grid value of $|\dot{h}'^{(i+1)}_{m,z} - \dot{h}'^{(i)}_{m,z}|$ becomes less than $10^{-3}$. The fictitious resistivity $\rho'_{out} = 2 \cdot 10^5$ was found sufficient to suppress the stray currents in vicinity of film boundaries. Our simulation results are presented in Table I.

Table I. Computation results

| $N$ of films | $Q_{mh}$, mJ | Computation time, hours |
|---|---|---|
| 4 | 3.43 | 9 |
| 6 | 3.51 | 23 |

Computed AC loss values coincide within 1-2% with the estimates in other works. Although we did not use the Parallel Computing Toolbox of Matlab, the employed Matlab FFT programs are intrinsically parallelized. Our computations were faster than those in [19]. Further numerical experiments showed that increasing the grid resolution, the resistivity $\rho_{out}$, and/or the computation domain size does not lead to $Q_{mh}$ variations exceeding 2% but the computation time grows quickly. In figure 1 we present the distribution of current density in the anisotropic bulk superconductor computed for $z' = z'_m$ as $\boldsymbol{J} = \boldsymbol{j}_m / d$; this distribution is very similar to those in [19, 20].

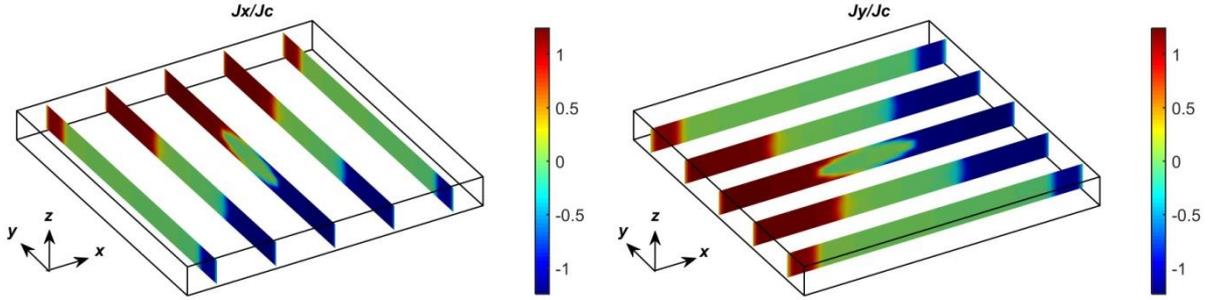

**Figure 1**. Current density at the peak of applied field: solution of the anisotropic bulk benchmark problem computed using the 6-film stack approximation.

We note that for the FFT-based stack simulations we needed to choose a larger computation domain $D$ than in the single film case (see, e.g., [21, 23]). As the distance between films decreases, the iterations (8)-(9) converge slower and also the evolutionary problem becomes stiff. Our results confirm that in such a case rescaling and solving the problem for an equivalent stack of only a few films can provide an efficient and accurate approximation.

## 4. Infinite stacks

Superconducting film stacks of a height several times greater the film size were used, e.g., in [7, 8]. For such, and also much shorter stacks, the current density distributions are, typically, similar in all films except the films close to the stack ends (see, e.g., the 2D solutions in [16]). Hence we will now consider an infinite stack of films, $\{(x, y, z_m) : (x, y) \in \Omega, z_m = md\}$, where $m$ takes all integer values. Here $\Omega$ can be an arbitrary finite 2D domain but, for simplicity, we assume again that $\Omega$ is simply connected.

All films in an infinite stack are under the same conditions, so for all films $g_m = g, h_{m,z} = h_z, j_m = j, e_m = e$. Since $A_{m,l}(k) = q^{|m-l|}$, where $q(k) = e^{-kd}$, equations (4) take the form

$$F[g] \sum_{l=-\infty}^{\infty} q^{|l|} = \frac{2}{k} F[h_z - h_{e,z}].$$

Summing the series up and taking the inverse Fourier transform we obtain

$$g = \Phi_\infty [h_z - h_{e,z}] := F^{-1}\left[ \frac{2(1-q)}{k(1+q)} F[h_z - h_{e,z}] \right] - C, \tag{11}$$

where, as was done above, the undetermined value of

$$S(k) = \frac{2(1-q)}{k(1+q)}$$

at $k=0$ is replaced by zero and $C$ chosen using the condition $\int_{\Omega_{out}} g \mathrm{d}\mathbf{r} = 0$. The evolutionary problem for $g$ is similar to that for a single film, see [23], with $S(k)$ instead of $2/k$. Briefly, the numerical method is as follows. For $g$ given at time $t$ we find $\mathbf{j} = \bar{\nabla} \times g$, set

$$\mathbf{e} = \begin{cases} \mathbf{e}(\mathbf{j}) & \text{in } \Omega, \\ \rho_{out} \mathbf{j} & \text{in } \Omega_{out}, \end{cases}$$

and use the Faraday law to define the initial approximation, $\dot{h}_z^{(0)} = -\mu_0^{-1} \nabla \times \mathbf{e}$; then shift it in $\Omega_{out}$, $\dot{h}_z^{(0)}|_{\Omega_{out}} := \dot{h}_z^{(0)}|_{\Omega_{out}} - A^{(0)}$, to satisfy the condition

$$\int_{R^2} \left( \dot{h}_z - \dot{h}_{e,z} \right) \mathrm{d}\mathbf{r} = 0. \tag{12}$$

We set $\dot{g}^{(0)} = \Phi_\infty(\dot{h}_z^{(0)} - \dot{h}_{e,z})$. The iterations

$$\begin{aligned}
\dot{h}_z^{(i+1)}\Big|_{\Omega_{out}} &= \dot{h}_z^{(i)}\Big|_{\Omega_{out}} - \lambda F^{-1}\left[\{S(k)\}^{-1} F[\dot{g}_{out}^{(i)}]\right]\Big|_{\Omega_{out}} - A^{(i)}, \\
\dot{g}^{(i+1)} &= \Phi_\infty[\dot{h}_z^{(i+1)} - \dot{h}_{e,z}],
\end{aligned} \tag{13}$$

are performed, for the spatially discretized problem, on each step of the ordinary differential equation solver until their convergence with a given tolerance. Here $\lambda$ is the relaxation parameter (as in [23], we used $\lambda = 0.7$), $\dot{g}_{out}^{(i)}$ is equal to zero in $\Omega$ and to $\dot{g}^{(i)}$ in $\Omega_{out}$, the shifts $A^{(i)}$ ensure that (12) holds on each iteration, and $\{S(k)\}^{-1} = k(1+q)/[2(1-q)]$ is set to zero at $k=0$, where also $1-q(0)=0$. Finally, the magnetic field can be expressed from (11) as $h_z = h_{e,z} + F^{-1}\left[\{S(k)\}^{-1} F[g]\right]$.

As an example, we simulated magnetization of the infinite stack of thin disks of radius $R$ characterized by the power law (1) with $n=50$. We used the dimensionless variables

$$(x', y', z') = \frac{(x, y, z)}{R}, \quad t' = \frac{t}{t_0}, \quad \mathbf{e}' = \frac{\mathbf{e}}{e_c},$$

$$\mathbf{j}' = \frac{\mathbf{j}}{j_c}, \quad \mathbf{h}' = \frac{\mathbf{h}}{j_c}, \quad g' = \frac{g}{j_c R},$$

where $t_0 = \mu_0 j_c R / e_c$, assumed the virgin initial state and the applied field $h'_{e,z} = t'$. Our simulations (figure 2) were for the film domain $\Omega' = \{r' < 1\}$ and the $1024 \times 1024$ grid in the computation domain $D' = \{(x', y') : -8 \le x', y' \le 8\}$; the iterations (13) were stopped when the average grid value of $|\dot{h}_z^{(i+1)} - \dot{h}_z^{(i)}|$ was less than $10^{-5}$. Numerical solution took from 20 minutes for $d=1$ to about 2 hours for $d=0.05$.

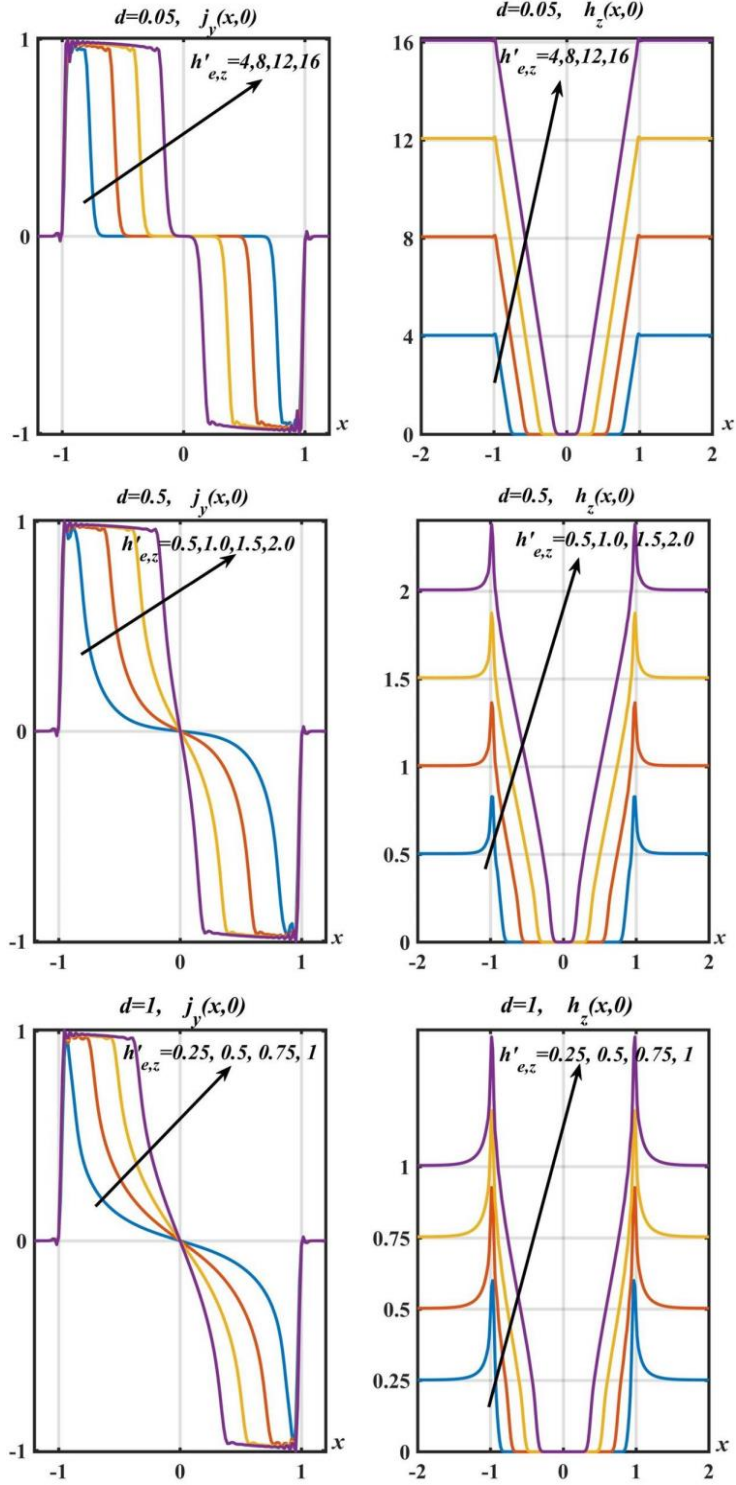

Figure 2. Magnetization of an infinite stack of thin disks: the distributions of $j_y$ (left) and $h_z$ (right) at $y=0$. The distance between disks, from top to bottom: 0.05, 0.5, 1 (dimensionless units).

We note that if an infinite stack is densely packed (figure 2, top), the solution is close to that for an infinite cylinder in a parallel field. As the distance between films increases, the stack problem solution tends to the solution for a single film (figure 2, bottom).

An interesting open question is whether solution to this specific infinite stack problem can, for the Bean model, be presented as a $d$-dependent algebraic transformation of the analytical solution [28] for thin disks (similarly to the case of strips, see [17]). On the other hand, the proposed numerical method is general: it is applicable to an infinite stack of arbitrary shaped films.

## 5. Discussion

Efficient numerical solution of highly nonlinear 3D eddy current problems in type-II superconductivity is necessary for design of superconducting devices. Several finite element methods have been proposed recently; some of them (see [19, 20]) were applied to 3D magnetization problems for an anisotropic bulk superconductor replacing a dense stack of thin superconducting films (the homogenized model).

Our work presents a simple but, nevertheless, efficient FFT-based alternative to these methods. Developed first for 2D film magnetization problems in [21, 22], the method was modified and extended to 3D bulk magnetization problems in our works [23, 27]. Here we adapted the FFT-based method to stacks of superconducting films, a perspective replacement of bulk superconductors in many practical applications.

We showed that, to simulate magnetization of a densely packed stack of a large number of films, transition to the anisotropic bulk model is not the only possible approach. As in the 2D case (stacks of infinitely long strips, see [16]), an accurate approximation can often be obtained as a solution to the magnetization problem for a stack of just a few films with the properly chosen characteristics. Replacing the anisotropic bulk superconductor by a stack of only four or six films, we computed an accurate solution to the stack benchmark problem: the obtained current density distributions were similar to those in [19, 20] and the AC loss estimates coincided within 1-2%.

The Matlab FFT software employed in our work is intrinsically parallelized; further parallelization of our algorithm is possible but was not implemented. Ran on a personal computer with four processors, our program was faster than those in [19].

The distributions of current in films of a high stack are usually similar in all except the films close to the stack ends. Previously, 2D magnetization problem for an infinite stack of long strips was solved analytically in [17]. Here we derived, for an infinite stack of arbitrary shaped flat films, a formulation and an FFT-based numerical method similar to those for a single film. Using the stack of thin disks as an example we illustrated the typical behavior of solution to such problems.

Although for simplicity we assumed that the films are simply connected, the multiply connected film case can be treated exactly as in the single film case (see, e.g., [23]).

**References**

[1] A. Patel, K. Filar, V. I. Nizhankovskii, S. C. Hopkins, and B. A. Glowacki, "Trapped fields greater than 7 T in a 12 mm square stack of commercial high-temperature superconducting tape," *Applied Physics Letters,* vol. 102, p. 102601, 2013.

[2] S. Zou, V. M. Zermeño, A. Baskys, A. Patel, F. Grilli, and B. Glowacki, "Simulation and experiments of stacks of high temperature superconducting coated conductors magnetized

by pulsed field magnetization with multi-pulse technique," *Superconductor Science and Technology,* vol. 30, p. 014010, 2016.

[3]   S. Zou, *Magnetization of high temperature superconducting trapped-field magnets* vol. 19: KIT Scientific Publishing, 2017.

[4]   A. Patel, S. Hahn, J. Voccio, A. Baskys, S. C. Hopkins, and B. A. Glowacki, "Magnetic levitation using a stack of high temperature superconducting tape annuli," *Superconductor Science and Technology,* vol. 30, p. 024007, 2016.

[5]   S. Kim, T. Kimoto, S. Hahn, Y. Iwasa, J. Voccio, and M. Tomita, "Study on optimization of YBCO thin film stack for compact NMR magnets," *Physica C: Superconductivity,* vol. 484, pp. 295-299, 2013.

[6]   K. Liu, W. Yang, G. Ma, L. Quéval, T. Gong, C. Ye*, et al.*, "Experiment and simulation of superconducting magnetic levitation with REBCO coated conductor stacks," *Superconductor Science and Technology,* vol. 31, p. 015013, 2017.

[7]   S. Hahn, Y. Kim, J. P. Voccio, J. Song, J. Bascunan, M. Tomita*, et al.*, "Temporal enhancement of trapped field in a compact NMR magnet comprising YBCO annuli," *IEEE Transactions on Applied Superconductivity,* vol. 24, pp. 1-5, 2014.

[8]   S. Hahn, J. Voccio, D. K. Park, K.-M. Kim, M. Tomita, J. Bascunan*, et al.*, "A stack of YBCO annuli, thin plate and bulk, for micro-NMR spectroscopy," *IEEE Transactions on Applied Superconductivity,* vol. 22, pp. 4302204-4302204, 2012.

[9]   M. Baghdadi, H. S. Ruiz, and T. A. Coombs, "Nature of the low magnetization decay on stacks of second generation superconducting tapes under crossed and rotating magnetic field experiments," *Scientific reports,* vol. 8, p. 1342, 2018.

[10]  M. Baghdadi, H. S. Ruiz, J.-F. Fagnard, M. Zhang, W. Wang, and T. A. Coombs, "Investigation of demagnetization in HTS stacked tapes implemented in electric machines as a result of crossed magnetic field," *IEEE Transactions on Applied Superconductivity,* vol. 25, pp. 1-4, 2015.

[11]  F. Grilli, S. P. Ashworth, and S. Stavrev, "Magnetization AC losses of stacks of YBCO coated conductors," *Physica C: Superconductivity,* vol. 434, pp. 185-190, 2006/02/15/ 2006.

[12]  E. Pardo, A. Sanchez, and C. Navau, "Magnetic properties of arrays of superconducting strips in a perpendicular field," *Physical Review B,* vol. 67, p. 104517, 03/31/ 2003.

[13]  R. Brambilla, F. Grilli, D. N. Nguyen, L. Martini, and F. Sirois, "AC losses in thin superconductors: the integral equation method applied to stacks and windings," *Superconductor Science and Technology,* vol. 22, p. 075018, 2009.

[14]  J. R. Clem, J. H. Claassen, and Y. Mawatari, "AC losses in a finite Z stack using an anisotropic homogeneous-medium approximation," *Superconductor Science and Technology,* vol. 20, p. 1130, 2007.

[15]  Y. Weijia, A. M. Campbell, and T. A. Coombs, "A model for calculating the AC losses of second-generation high temperature superconductor pancake coils," *Superconductor Science and Technology,* vol. 22, p. 075028, 2009.

[16]  L. Prigozhin and V. Sokolovsky, "Computing AC losses in stacks of high-temperature superconducting tapes," *Superconductor Science and Technology,* vol. 24, p. 075012, 2011.

[17]  Y. Mawatari, "Critical state of periodically arranged superconducting-strip lines in perpendicular fields," *Physical Review B,* vol. 54, pp. 13215-13221, 11/01/ 1996.

[18]  E. H. Brandt and M. Indenbom, "Type-II-superconductor strip with current in a perpendicular magnetic field," *Physical review B,* vol. 48, p. 12893, 1993.

[19]  M. Kapolka, V. M. R. Zermeño, S. Zou, A. Morandi, P. L. Ribani, E. Pardo*, et al.*, "Three-Dimensional Modeling of the Magnetization of Superconducting Rectangular-Based Bulks and Tape Stacks," *IEEE Transactions on Applied Superconductivity,* vol. 28, pp. 1-6, 2018.

[20]  M. Olm, S. Badia, and A. F. Martín, "Simulation of high temperature superconductors and experimental validation," *arXiv preprint arXiv:1707.09783,* 2017.

[21] J. I. Vestgården and T. H. Johansen, "Modeling non-local electrodynamics in superconducting films: the case of a right angle corner," *Superconductor Science and Technology,* vol. 25, p. 104001, 2012.

[22] J. I. Vestgården, P. Mikheenko, Y. M. Galperin, and T. H. Johansen, "Nonlocal electrodynamics of normal and superconducting films," *New Journal of Physics,* vol. 15, p. 093001, 2013.

[23] L. Prigozhin and V. Sokolovsky, "Fast Fourier transform-based solution of 2D and 3D magnetization problems in type-II superconductivity," *Superconductor Science and Technology,* vol. 31, p. 055018, 2018.

[24] D. Champeney, *Fourier Transforms and Their Physical Applications*. New York: Academic, 1973.

[25] M. Abramowitz and I. A. Stegun, *Handbook of mathematical functions: with formulas, graphs, and mathematical tables* vol. 55: Courier Corporation, 1964.

[26] E. Pardo and M. Kapolka, "3D computation of non-linear eddy currents: Variational method and superconducting cubic bulk," *Journal of Computational Physics,* vol. 344, pp. 339-363, 2017/09/01/ 2017.

[27] L. Prigozhin and V. Sokolovsky, "3D simulation of superconducting magnetic shields and lenses using the fast Fourier transform," *Journal of Applied Physics,* vol. 123, p. 233901, 2018.

[28] P. Mikheenko and Y. E. Kuzovlev, "Inductance measurements of HTSC films with high critical currents," *Physica C: Superconductivity,* vol. 204, pp. 229-236, 1993.